\documentclass[twocolumn,amsmath,amssymb,prb,superscriptaddress,longbibliography,nobalancelastpage]{revtex4-1}

\usepackage{titlesec}
\usepackage[colorlinks,bookmarks=false,citecolor=blue,linkcolor=blue,urlcolor=blue]{hyperref}
\usepackage{epsfig}
\usepackage{color}
\usepackage{mathtools}
\usepackage{subfigure}
\usepackage{multirow}
\usepackage{rotating}
\usepackage{lipsum}
\usepackage{graphicx} 
\usepackage{dcolumn}  
\usepackage{amsmath,bm}

\input{insbox}
\usepackage[titletoc,title]{appendix}
\usepackage[graphicx]{realboxes}
\usepackage{graphicx,subfigure}

        {\pagebreak[4]\global\pdfpageattr\expandafter{\the\pdfpageattr/Rotate 0}}%

\newcommand{\be}{\begin{equation}}
\newcommand{\ee}{\end{equation}}

\newcommand{\CFO}{CoFe$_2$O$_4$ }
\newcommand{\FO}{Fe$_3$O$_4$ }

    \definecolor{treegreen}{RGB}{51, 102, 0}
    
    \definecolor{navy}{RGB}{0, 0, 128}

\begin{document}

\title{Macrospin model of an assembly of magnetically coupled core-shell nanoparticles}

\author{Nikolaos Ntallis}
\affiliation{Department of Physics and Astronomy, Uppsala University, Uppsala 751 20, Sweden}

\author{Corisa Kons}
\affiliation{Department of Physics, University of South Florida, Tampa, Florida 33620, USA}

\author{Hariharan Srikanth}
\affiliation{Department of Physics, University of South Florida, Tampa, Florida 33620, USA}

\author{Manh-Huong Phan}
\affiliation{Department of Physics, University of South Florida, Tampa, Florida 33620, USA}

\author{D.A. Arena}
\affiliation{Department of Physics, University of South Florida, Tampa, Florida 33620, USA}

\author{Manuel Pereiro}
\email[Corresponding author: ]{manuel.pereiro@physics.uu.se}
\affiliation{Department of Physics and Astronomy, Uppsala University, Uppsala 751 20, Sweden}

\begin{abstract}


Highly sophisticated synthesis methods and experimental techniques allow for precise measurements of magnetic properties of nanoparticles that can be reliably reproduced using theoretical models. Here, we investigate the magnetic properties of ferrite nanoparticles by using theoretical techniques based on Monte Carlo methods. We introduce three stages of sophistication in the macromagnetic model. First, by using tailor-made hamiltonians we study single nanoparticles. In a second stage, the internal structure of the nanoparticle is taken into consideration by defining an internal (core) and external (shell) region, respectively. In the last stage, an assembly of core/shell NPs are considered. All internal magnetic couplings such as inter and intra-atomic exchange interactions or magnetocrystalline anisotropies have been estimated. Moreover, the hysteresis loops of the aforementioned three cases have been calculated and compared with recent experimental measurements. In the case of the assembly of nanoparticles, the hysteresis loops together with the zero-field cooling and field cooling curves are shown to be in a very good agreement with the experimental data. The current model provides an important tool to understand the internal structure of the nanoparticles together with the complex internal spin interactions of the core-shell ferrite nanoparticles.

\end{abstract}

\maketitle

\section{Introduction}
Various technological fields have motivated the design and fabrication of nanostructures with the ability of tailoring the magnetic properties. The category of magnetic nanoparticles (NPs) are interesting from both fundamental and technological points of view \cite{par1,par2,par3,par4,par5,par6}. Thus, for example, ferrite nanoparticles with the general formula of MFe$_2$O$_4$ (M = Fe, Co, Ni and Mn) have attracted great attention of researchers due to their potential applications in biomedicine and industry \cite{par2}. In particular, spinel ferrites are an attractive class of ferrimagnets offering both soft and hard phases as well as a common crystal structure that allow for a high quality crystal interface between core and shell layers \cite{mag_NPs}. The metallic ions are located at either octahedral  coordinated sites---also called B-sites---or forming a tetrahedral geometry---denoted as A-sites. The ratio of the occupation of these sites produces different spinel structures defined by the inversion degree number $X$. The normal  and inverse spinel structures represent the two limiting cases with $X=0$ and $X=1$, respectively. Between these two limits, a mixed spinel exists where the divalent transition metal M is distributed between sites A and  B~\cite{coey,cullity}. A value of $ X=2/3$ represents a random distribution of metallic ions between sites A and B. The degree of inversion can affect the magnetic properties of ferrites as, for example, saturation magnetization and coercive field ~\cite{an1,an2}. Ferrites are found to have a ferrimagnetic magnetic ordering, which is due to the dominant antiferromagnetic  exchange between  sites A and B~\cite{coey,cullity}. 

The inverse spinel structure is of paramount importance for the current work where Fe$^{3+}$ cations are equally distributed at both A and B sites while the divalent M ions are found only at octahedral sites. In this work, \CFO (CFO) was selected due to its high anisotropy that should limit the degree of canting as spins will be more tightly bound to the crystal lattice compared to \FO (FO), which has a relatively high moment but much lower anisotropy. A common crystal structure and negligible differences in lattice constant between the two materials (8.40~\AA~for FO, 8.39~\AA~for CFO \cite{coey}) enables synthesis of high quality core/shell NPs and, consequently, a bi-magnetic structure of these two compounds can be constructed without introducing  large lattice mismatch distortions. Moreover, as CFO is in a hard magnetic phase while FO is in a soft magnetic phase at normal conditions, the variants of these compounds can possess interesting exchanged coupled related properties. 

Thus, two variants of core/shell NP assemblies have already been synthesized and magnetically characterized \cite{experiment}. More specifically, a conventional assembly of NPs is experimentally synthesised by adding CFO to the core of the NP and capping the core with a shell of FO (CFO@FO). The inverted assembly is achieved by placing FO in the core and adding CFO in the shell (FO@CFO). In order to give a better understanding of their underlying reversal mechanisms, we employ Monte Carlo (MC) atomistic simulations for both type of assemblies based on tailor-made hamiltonians. Thus, we introduce progressively different degrees of sophistication of the model. Initially, we performed MC simulations for single nanoparticles by varying several of their internal degrees of freedom in order to examine how interface and surface effects affect their magnetic response. In a second stage, the hamiltonian is improved to describe a single nanoparticle composed of a central region denoted as core and an external shell region. The hamiltonian also considers the interface effects between the core and shell regions. Finally, the hamiltonian, and consequently, the macrospin model is recasted in a more sophisticated shape so that the model now is capable to describe not only single nanoparticles with internal structure but also an assembly of core-shell nanoparticles. A depiction of the three NP systems modeled is shown in Fig.~\ref{cartoon}. Moreover, we also indicate in several colors the regions of the nanoparticles that are essential to understand the different terms considered in the hamiltonians given by Eqs.~(\ref{model}-\ref{model_assembly}).

\section{Model and Results}
\begin{figure}
\includegraphics[width=8.5cm]{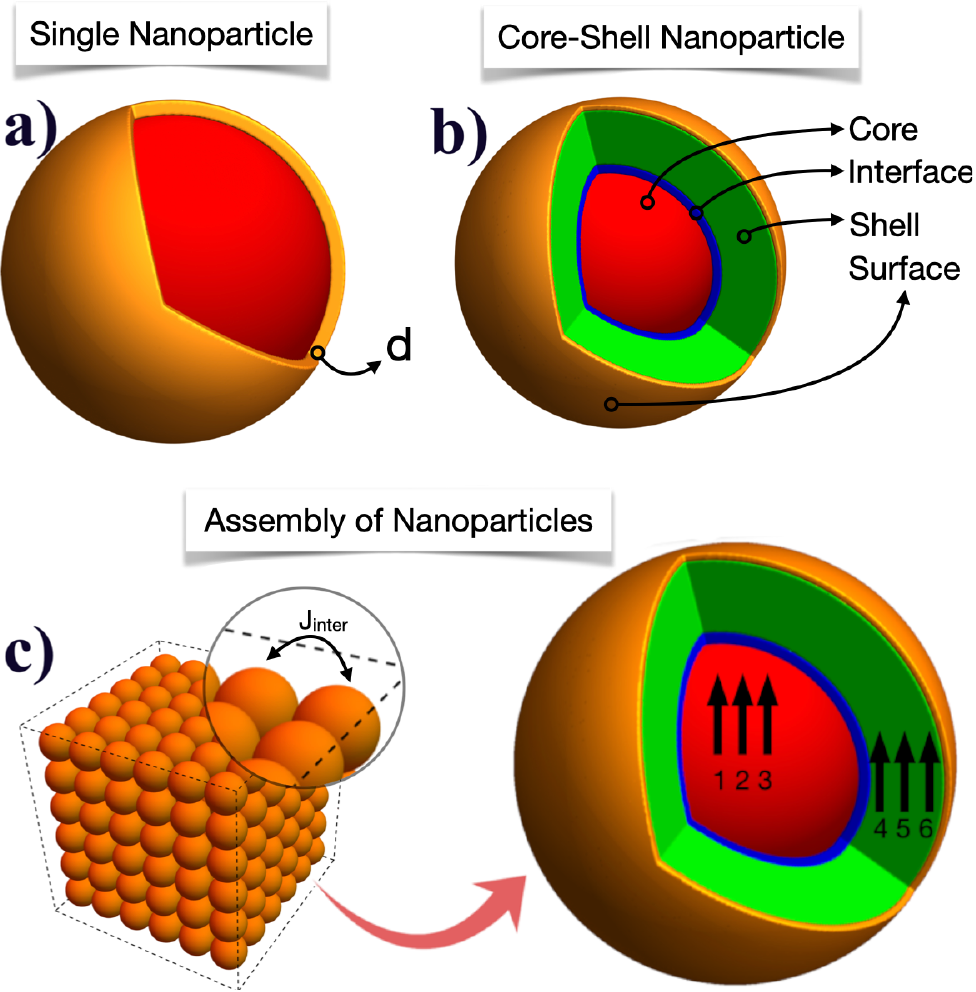}
\caption{Since the studied nanoparticles are very small in size, the surface effects become very important. Thus, we sketch in a) an spherical single nanoparticle with the region in gold representing the portion of the nanoparticle that belongs to the surface. The width of the surface is denoted by d.  In b), a core-shell nanoparticle is shown with the different regions indicated by using different colors. Finally, in c) is represented an assembly of core-shell nanoparticles, while on the right is drawn the internal structure of a single nanoparticle taken from the assembly. The arrows represent the macrospins used in the model. 
}
\label{cartoon}
\end{figure}

Initally, we employed atomistic Monte Carlo simulations with the implementation of the Metropolis algorithm of isolated nanoparticles composed of \CFO and Fe$_3$O$_4$. The Hamiltonian used for the calculations is 
\begin{equation}
	\mathcal{H}=-\sum_{i,j}J_{ij} \bm{S}_i \bm{S}_j-\sum_{i}K_{i}\cos^2\bm{\theta}_i -g\mu_B\sum_i \bm{B} \bm{S}_i 
	\label{model}
\end{equation}
\noindent where $i, j$ denote first-neighbour atomic positions and $\bm{S}_i$ is the atomic magnetic moment. The first term describes the Heisenberg exchange interaction, $J_{ij}$, between atomic sites $i$ and $j$. The ansatz given in Eq.~(\ref{model}) implies that a ferromagnetic (antiferromagnetic) interaction is described by a positive (negative) exchange coupling. In the second term, a uniaxial atomic magnetic anisotropy of strength $K_{i}$ is introduced to collect the spin-orbit effects. The angle ${\theta}_{i}$ defines the deviation of the atomic magnetic moment ($\bm{S}_i$) with respect to the atomic easy axis anisotropy. The last term represents the Zeeman term under the influence of a global external magnetic field. For the very small sizes in diameter of the nanoparticles considered here ($\sim 6$ nm for CFO and $\sim 7$ nm for FO), the surface effects become  important, and thus, for all the simulated systems, unless stated otherwise, we define a surface shell thickness of width $d$ (cf. Fig.~\ref{cartoon} a)). The width $d$ is chosen to be equal to the unit cell length of the associated material. For the core spins, the easy axis anisotropy is set along $z$ direction whereas for the surface atoms the anisotropy axis direction is randomly selected for each spin site in agreement with experimental evidence \cite{experiment}. Bulk $J_{ij}$ values are collected in Table~\ref{tab:CFO+FO_Jij} for CFO and FO nanoparticles \cite{Jij_s}. Likewise, the bulk values of the magnetic anisotropy for the core region (red color in Fig.~\ref{cartoon} a)) of the nanoparticles are $k_{i}=0.0036$~mRy and $k_{i}=0.000112$~mRy per atom for CFO and FO, respectively~\cite{coey,cullity}. Both CFO and FO have an inverse spinel structure, i.e., the divalent ion (Co$^{+2}$ for CFO and Fe$^{+2}$ for FO) is found only in B sites.  The Co$^{+2}$ ion has a moment of $3.0~\mu_{B}/$atom, whereas Fe$^{+2}$ presents a moment of $4.0~\mu_{B}/$atom. In both samples, Fe$^{+3}$ ions have a moment of 5.0~$\mu_{B}/$atom. The spherical particles with radius $r$ are constructed by replication of the bulk inverse spinel structure. The spherical shape is reproduced by removing the unit cells that are a distance bigger than $r$ away from center of the nanoparticle.

\begin{table}[t]
 \caption{Bulk exchange coupling constants for CFO and FO with different interaction transition-metal (TM) atomic sites. All values are given in energy units of mRy.}
\vspace{5pt}
 \centering
 \begin{ruledtabular}
 \begin{tabular}{lccccc}
      CFO & $TM_{i}$ & $TM_{j}$ & $i$ & $j$ & $J_{ij}$   \\
          & Fe  & Fe & A & A & $-0.094$ \\
          & Fe  & Co & A & B & $-0.143$ \\
          & Fe  & Fe & A & B & $-0.164$ \\
          & Co & Co & B & B & $0.296$ \\
          & Fe  & Co & B & B & $-0.117$ \\
          & Fe  & Fe & B & B & $-0.047$ \\
      \hline
      FO  & Fe  & Fe & A & A & $-0.132$ \\
          & Fe  & Fe & A & B & $-0.150$ \\
          & Fe  & Fe & A & B & $-0.177$ \\
          & Fe & Fe & B & B & $0.308$ \\
          & Fe  & Fe & B & B & $-0.08$ \\
          & Fe  & Fe & B & B & $-0.06$ \\
 \end{tabular}
 \end{ruledtabular}
    \label{tab:CFO+FO_Jij}
\end{table}

\begin{figure}
\includegraphics[width=8.5cm]{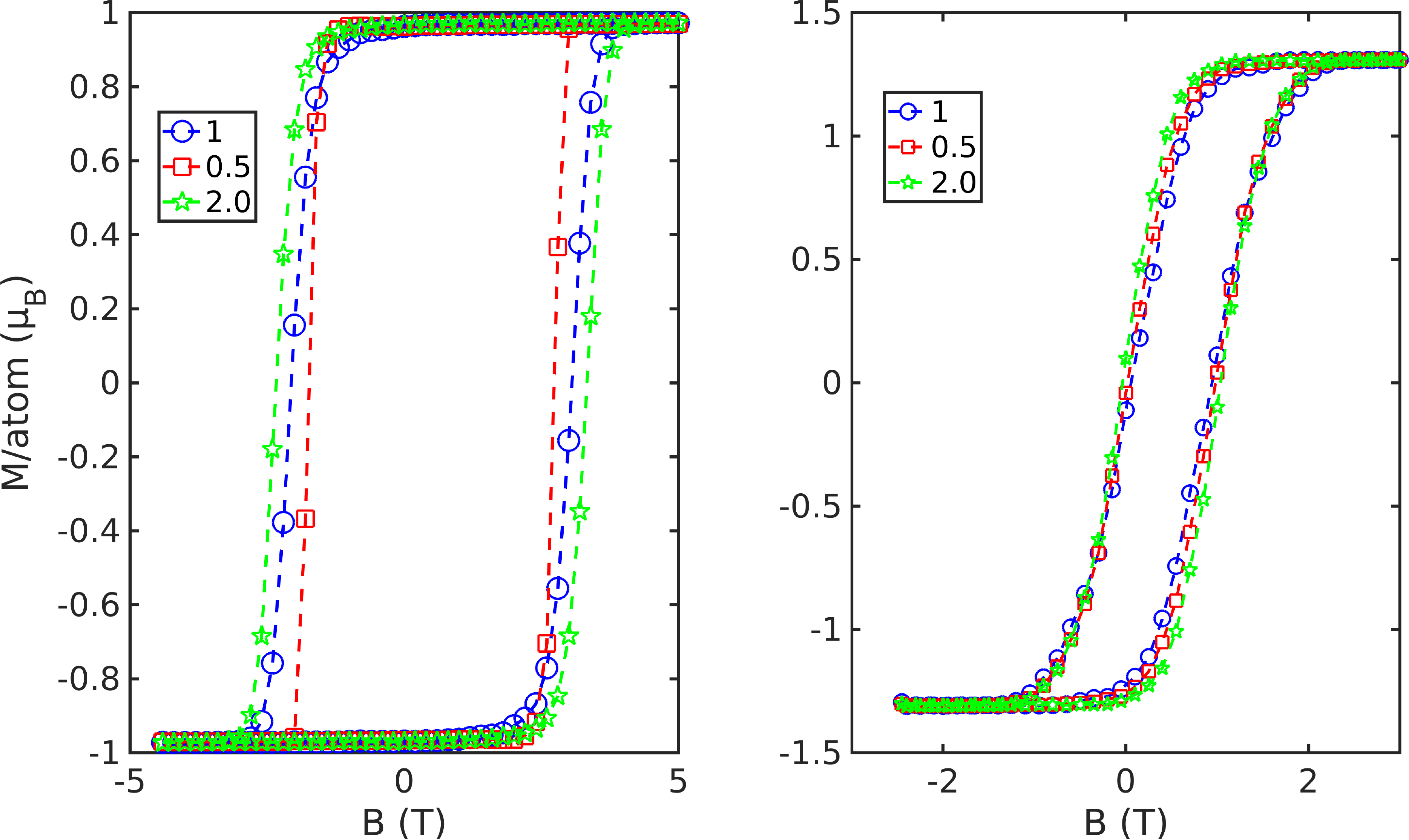}
\caption{Calculated hysteresis loops for CFO (left panel) and FO (right panel) single nanoparticles for different ratios of the surface anisotropy with respect to the core anisotropy $\left(\frac{K_i^s}{K_i^c}\right)$ at a temperature of 5 K.
}
\label{cfo-fo-loops5k}
\end{figure}

Figure~\ref{cfo-fo-loops5k} shows hysteresis loops for CFO and FO single nanoparticles at $T=5K$ by varying the strength of the surface anisotropy ($K_i^s$) with respect to the one of the core ($K_i^c$). As surface atoms interact with a number of neighbouring oxygen atoms smaller than the number of oxygen atoms interacting with bulk atoms, the exchange couplings of the surface atoms are expected to be smaller than the exchange couplings for bulk atoms. Thus, we set $J_{ij}^s/J_{ij}^c=0.5$. This value is chosen on a perfect spherical surface formed by a cubic unit cell with the mean value of the coordination number drooping to half in the surface. Thus, by setting the aforementioned ratio to $0.5$ we assume that, in mean average, the coordination number in the surface is half of the one for the atoms in the bulk. FO single nanoparticles clearly shows a larger saturation magnetization with a value of $1.31~\mu_{B}/$atom  with respect to CFO which has a value of $0.97~\mu_{B}/$atom. On the contrary, CFO possesses a larger coercive field reaching a value $\sim 2.5$ T. It is worthwhile to mention here that the  variation of the coercive field with respect to the ratio $K_i^s/K_i^c$ is almost negligible, in particular, for FO single nanoparticles. This is a direct consequence of the the randomly easy axis distribution of the surface atoms.


By increasing the level of sophistication of the model, we proceed to the calculation of single core-shell nanoparticles for the two aforementioned variants, i.e.,  CFO@FO and FO@CFO. In order to study the core-shell morphology, we introduce the following hamiltonian:

\begin{eqnarray}
	\mathcal{H}=&-&\sum_{i,j}^{\rm{bulk}} J_{ij} \bm{S}_i\bm{S}_j -\sum_{i,j}^{\rm{interf.}} a_{i} J_{ij} \bm{S}_i \bm{S}_j -\sum_{i,j}^{\rm{surf.}} a_{s} J_{ij} \bm{S}_i \bm{S}_j\nonumber\\
	&-&\sum_{i}K_{i}\cos^2\bm{\theta}_i -g\mu_B\sum_i \bm{B} \bm{S}_i 
	\label{model_coreshell}
\end{eqnarray}

In Eq.~(\ref{model_coreshell}), we have introduced two additional terms which separate the interface and surface effects from the bulk of the nanoparticle. We assume that both the width of the surface (d$_s$) and interface (d$_i$) are equal to the CFO unit cell constant, i.e. d$_s$=d$_i$=$0.835$ nm. As both CFO and FO possess the same inverse spinel structure and very similar lattice constant, we would expect an almost perfect match in the interface. However, due to the spherical shape of the nanoparticle, and consequenly, the curvature of the interface, there is still a slight mismatch between the boundaries in the interface which produces a nonzero surface tension. For the aforementioned reasons, in the current model we assume that the spinel structure is preserved but it is slightly distorted on the interface and surface. The distortion is introduced by rescaling the exchange constants $J_{ij}$ on the interface by the factor $a_{i}$ and on the surface by $a_{s}$. The initial value of $J_{ij}$ on the interface is assumed to be the mean average of the sum of $J_{ij}$ of the associated bulk values of the different structures. Unless stated elsewhere, $a_{s}$ has a value of $0.5$, i.e., on the surface we assume that interaction strengths drop to the half value than the one found in the bulk. As a consequence, we assume the anisotropy becomes randomized on the surface but uniaxial on the interface. The structural parameters of the core-shell nanoparticles studied here are given in Table~\ref{tab:size}.

\begin{table}[t]
 \caption{ Structural parameters of CFO@FO and FO@CFO nanoparticles. All values are given in distance units of nm.}
\vspace{5pt}
 \centering
 \begin{ruledtabular}
 \begin{tabular}{lccc}
             & Core Radius & Shell Thickness & Diameter   \\
    CFO@FO   & 2.9  &  1.7 & 9.2\\
    FO@CFO   & 3.5  &  1.3 & 9.6 \\
 \end{tabular}
 \end{ruledtabular}
    \label{tab:size}
\end{table}

\begin{figure}
\includegraphics[width=8.5cm]{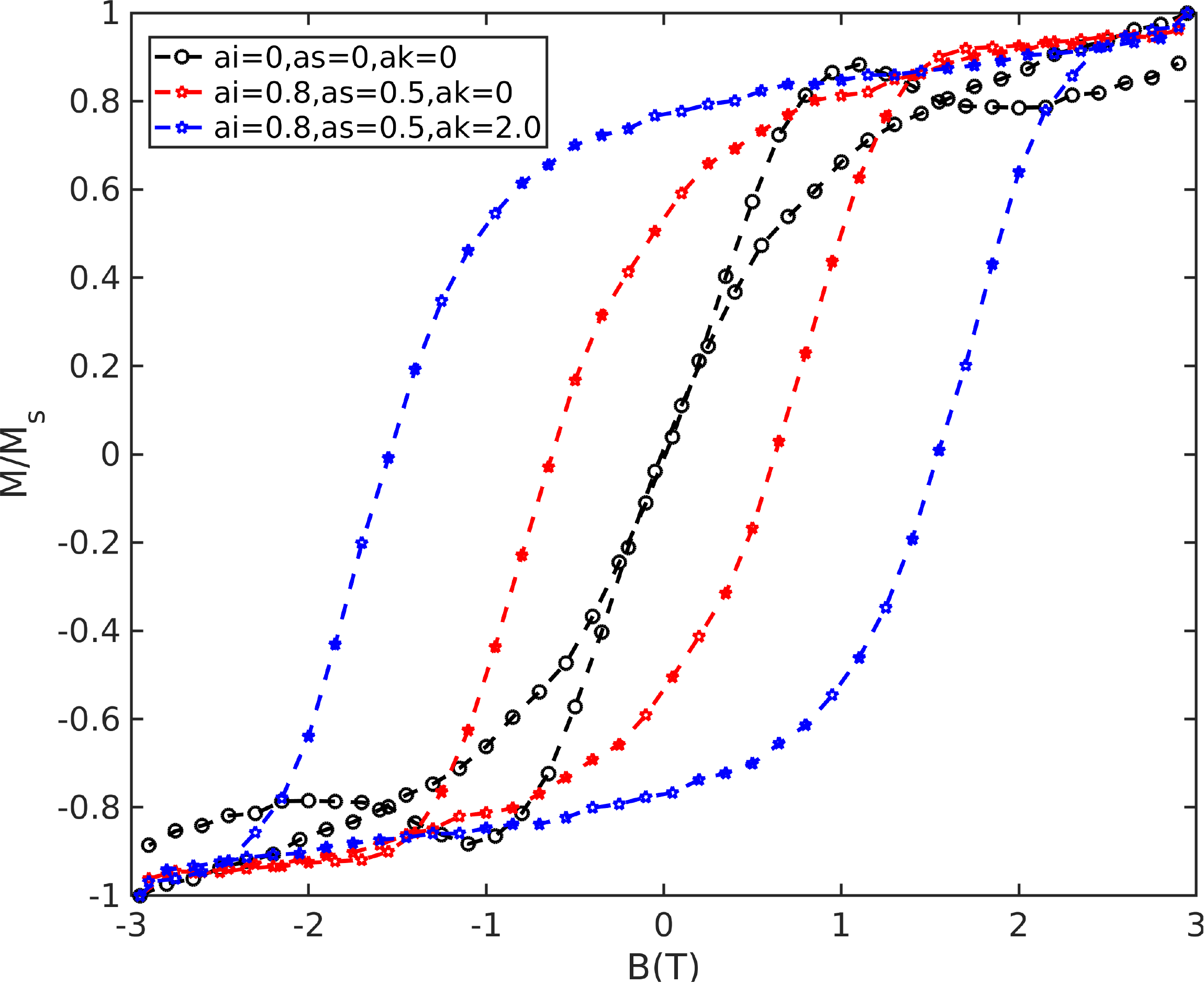}
\caption{Hysteresis loops for CFO@FO core-shell nanoparticle at 5 K. Loops are plotted for different scaling factors $a_{i}$, $a_{s}$ and $a_{k}$. 
}
\label{cfo-fo_nano}
\end{figure}

\begin{figure}
\includegraphics[width=8.5cm]{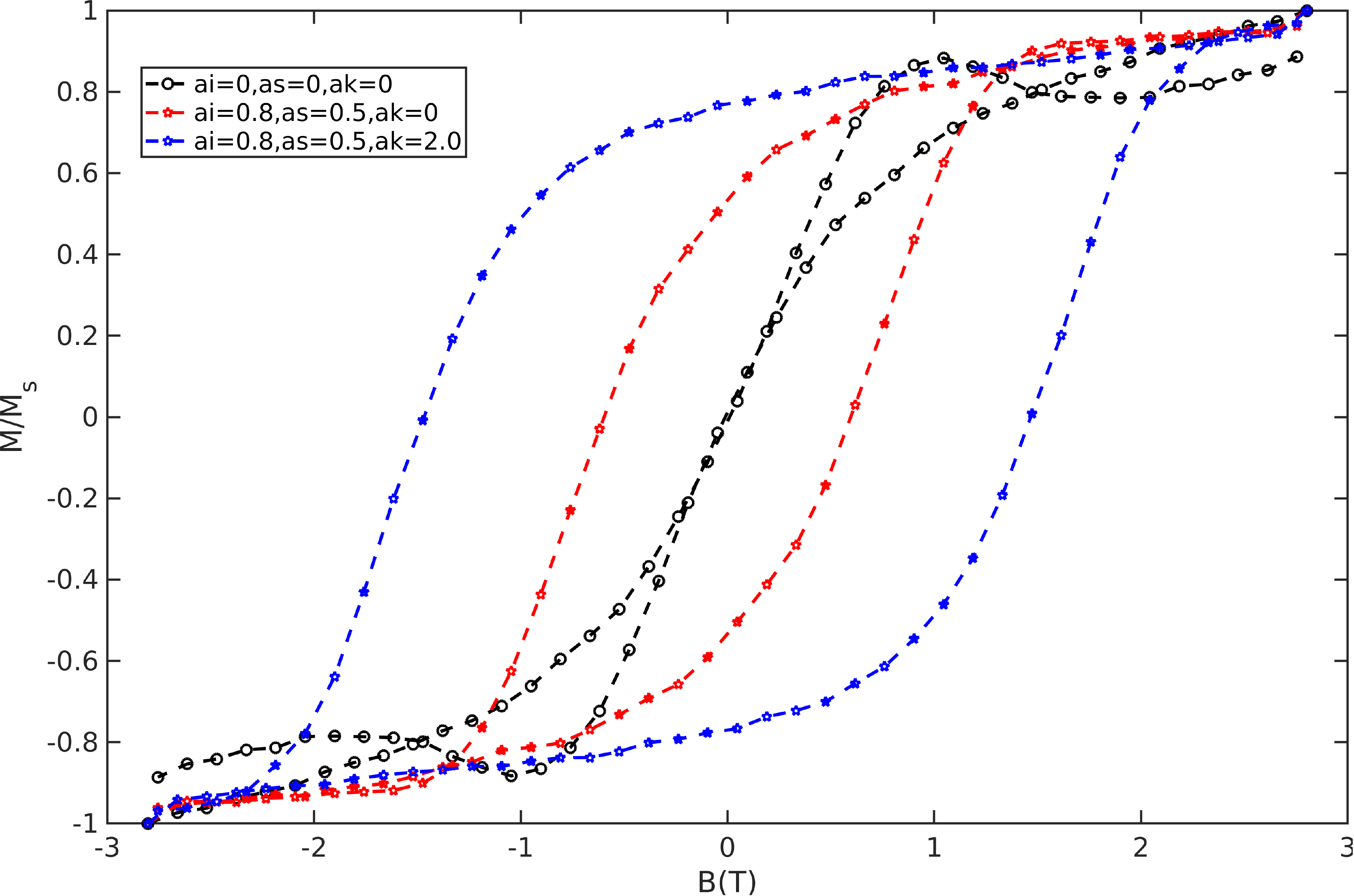}
\caption{Hysteresis loops for FO@CFO single nanoparticle at 5 K. Loops are plotted for different scaling factors $a_{i}$, $a_{s}$ and $a_{k}$. 
}
\label{fo-cfo_nano}
\end{figure}

It has to be noted here that the size of both types of core-shell nanoparticles are far below the coherent radius limit of ~$3.6 l_{ex}$, where $l_{ex}$ is the exchange length~\cite{ntallis}. Typical values of the exchange length are 4.9 nm and 5.2 nm for Fe$_3$O$_4$ and CoFe$_2$O$_4$ bulk systems, respectively~\cite{coey}. Thus, domain formation   is very unlikely. In consequence, the magnetization reversal process can only be attributed to incoherent states on the interface or surface. Figures ~\ref{cfo-fo_nano} and ~\ref{fo-cfo_nano} show hysteresis loops at 5K for CFO@FO and FO@CFO core-shell nanoparticles, respectively. The hystersis loops are plotted for different scaling factors, i.e., $a_i$, $a_s$ and $a_k$. The anisotropy scaling parameter per atom is represented by $a_{k}$. Both types of nanoparticles possess similar properties as single particles. Thus, under the absence of magnetocrystalline anisotropy (black dotted curves, $a_k=0$) both types of core-shell nanoparticles show a {\it knee}-like behaviour close to saturation indicating the frustration produced by $J_{ij}$ in ferrites. Also in the range from -$3$ T to $3$ T in the applied external magnetic field, none of the CFO@FO and FO@CFO core-shell nanoparticles possess a full closed loop. However, by scaling (reducing) the exchange constants on the interface without magnetocrystalline anisotropy, the aforementioned effect is highly reduced  although it is worthwhile to mention here that the core-shell nanoparticles still have a non-zero coercive field (cf. red curves in Figs.~\ref{cfo-fo_nano} and \ref{fo-cfo_nano}). The latter arises from the fact that the mismatch between the core and shell introduces canting in the interface spins and thus induces an increase of the exchange anisotropy of the system. Introduction of magnetocrystalline anisotropy into the nanoparticles led to an increase in the coercive field. For example, for a scaling factor $a_{k}=2.2$, the coercive field is calculated to be $2.3$ T and $1.9$ T for FO@CFO and  CFO@FO, respectively.

\begin{figure}
\includegraphics[width=8.5cm]{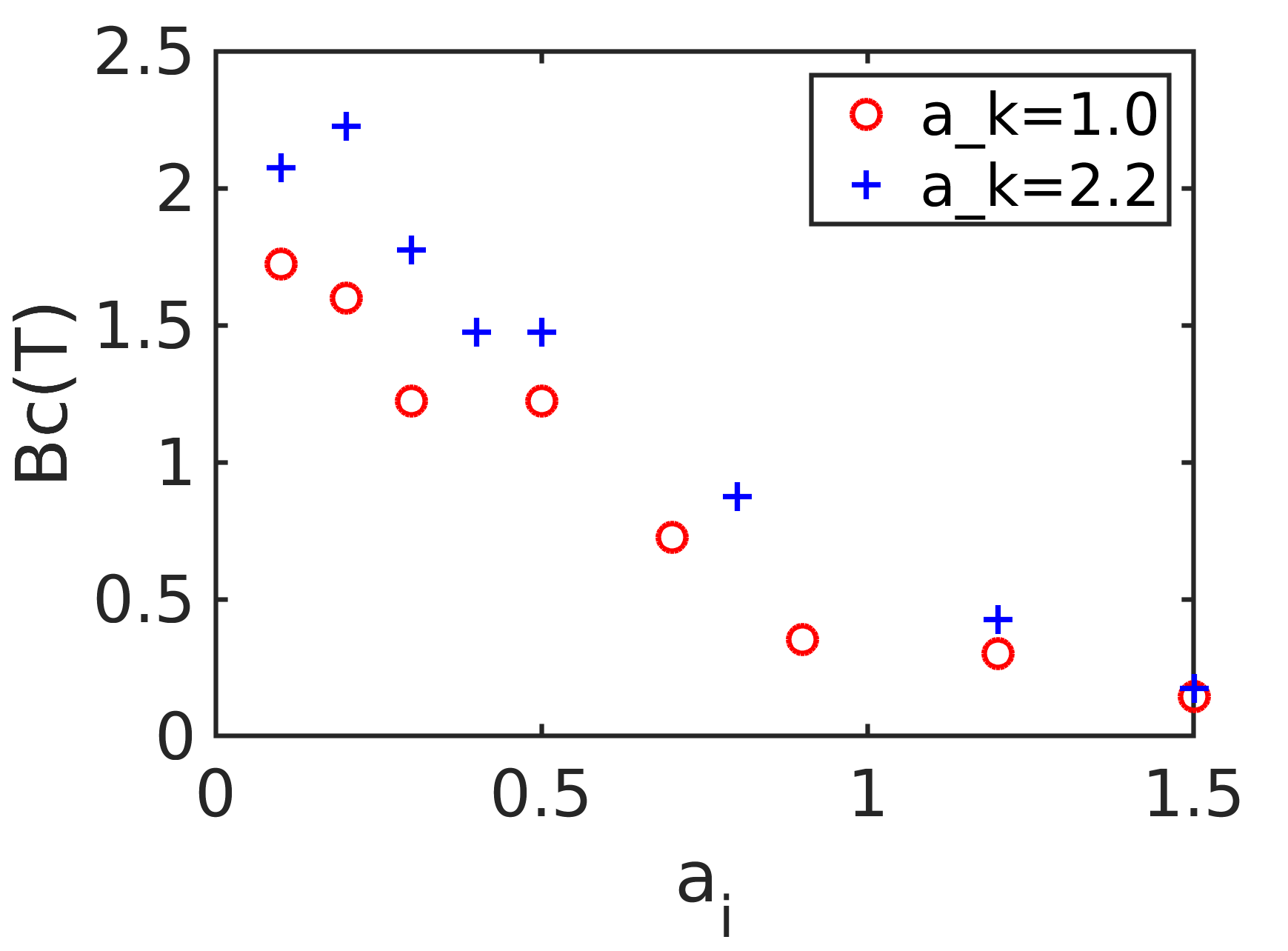}
\caption{Coercive field as a function of the interface scaling factor $a_{i}$ for CFO@FO nanoparticles.
}
\label{cfo-fo_coerce}
\end{figure}

For exchange coupled systems, the exchange interaction on the interface is a critical parameter regardless of the soft or hard phase of the nanoparticle. Figure ~\ref{cfo-fo_coerce} shows the variation of the coercive field with respect to the interface scaling factor for CFO@FO nanoparticle at 5K for two different values of the anisotropy scaling factor. In the limit of no anisotropy coupling ($a_{i}\rightarrow0$), the coercive field approximates to that of the hard phase. For both anisotropy scaling parameters, the coercive field possess a non-monotonic behaviour. When the anisotropy constant approaches  the bulk values, i.e., $a_{k}\rightarrow1$, the coercive field continuously decreases but the decay is broken in the interval $[0.3,0.5]$. In this region, the coercive field curve shows a plateau with a very tiny increase. After this value the decrease rate reduces, i.e., the coercive field decreases but with a smaller slope. On the other hand, for $a_{k}=2.2$ the coercive field shows a  peak for $a_{i}\sim0.2$, so that, the interface effects induce magnetic fluctuations for such coupled systems. Notably the hardness of the bi-magnetic systems is heavily dependent on the interplay between the anisotropy barrier and the exchange interactions on the interface. As already mentioned above, the disorder induced by the interface effects, even under the influence of  strong anisotropy, results in a non-zero coercive field. By increasing of the anisotropic scaling factor, a high coercive field can be reproduced with a small interface coupling factor. On the other hand, for the anisotropy bulk value case, a very strong interface coupling must be considered in order to achieve the maximum  coercive field. Therefore, a nanoparticle system requires an almost perfect interface with negligible imperfections or dislocations. From these facts, we can conclude that for bi-magnetic nanoparticle systems, the interface shows an enhanced anisotropy.

\begin{figure}
\includegraphics[width=8.5cm]{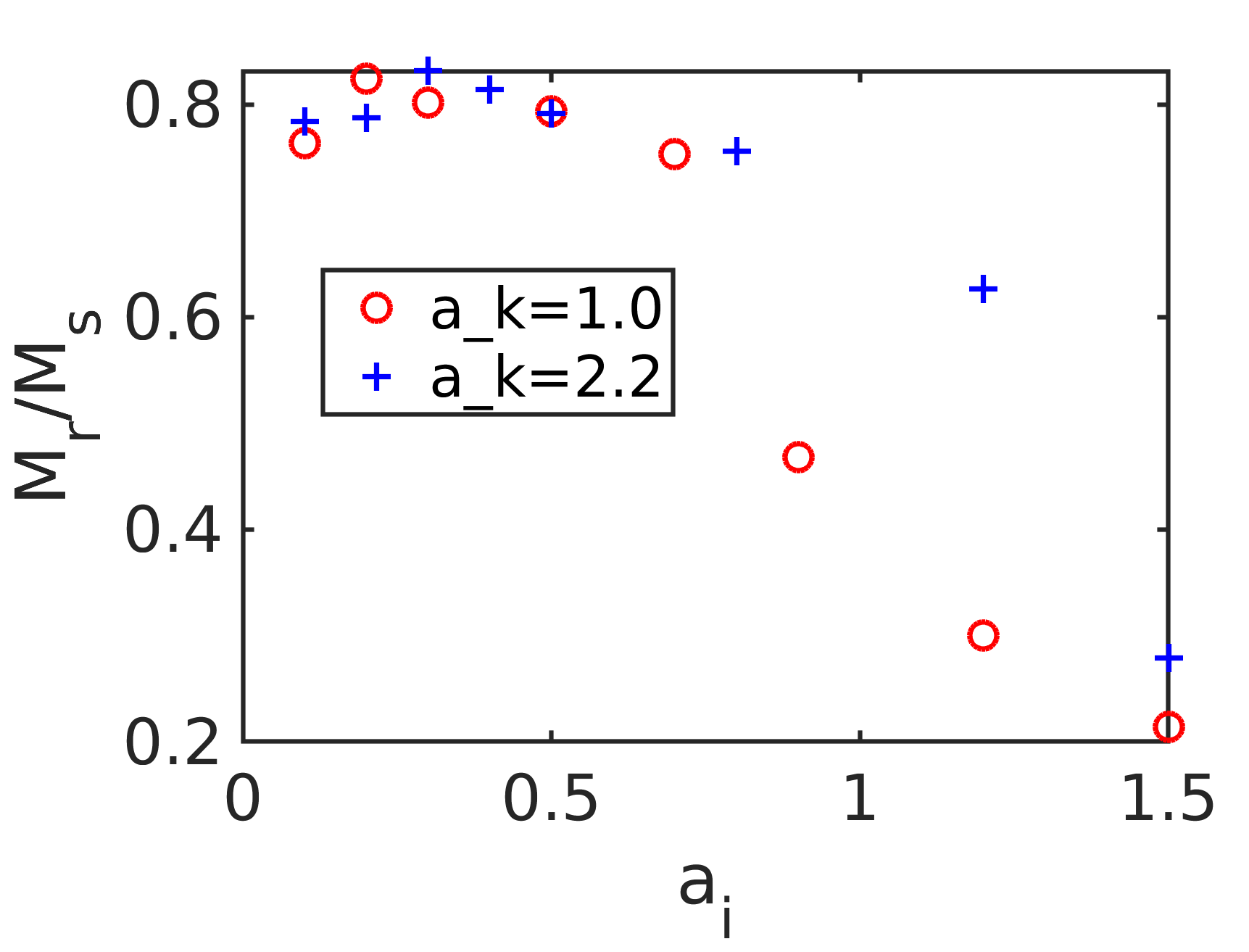}
\caption{Variance of the normalized remanent magnetization for CFO@FO nanoparticle with respect to interface scaling factor $a_{i}$. 
}
\label{cfo-fo_rema}
\end{figure}

Apart form the coercive field, the remanent magnetization is a critical parameter for magnetic applications. Figure ~\ref{cfo-fo_rema} shows the evolution of the normalized remanent magnetization with respect to the interface scaling factor. For both  $a_{k}=1.0$  and $a_{k}=2.2$, the remanent magnetization shows an increasing trend up to a critical value of the interface scaling factor $a_{i}=0.3$ and then clearly reduces down to a value of $M_r/M_s=0.2$. For both anistropy scaling factors, the $M_{r}/M_{s}$ ratio show a maximum close to $0.8$. 

 Independently from the interface coupling, the surface is free to start a reversal process due to reduced coordination of the atomic spins. From our simulations it is found that, in order to reach the maximum  $M_{r}/M_{s}$ a larger value of the interface coupling constant is needed, with respect to the one for the maximum coercive field. As the surface and interface are already disordered an even stronger coupling between the phases is needed in order for the soft phase to overrule and rise the remament state. Interestingly though when the $M_{r}/M_{s}$  is at its maximum value still the nanoparticle possess a considerable coercive field as the disorder of the interface does not allow for a full reversal process to be completed. Despite that, it is clear that in ferrites the exchange interactions are extremely strong  and play  a major role in the magnetic behaviour. As seen in Figs.~\ref{cfo-fo_coerce} and ~\ref{cfo-fo_rema}, even with a scaling of $a_{k}=2.2$ for a value of $a_{i}=0.3$, a clearly exchanged coupled behaviour can been achieved. Moreover, the size of the particle is also extremely important as it does not allow for any kind of domain formation~\cite{ntallis}.

Experimental measurements have been performed on dense assemblies of CFO@FO and FO@CFO core-shell nanoparticles~\cite{experiment}. Under these experimental conditions, the dipolar interaction can affect the magnetic response, especially for NPs composed of CFO and FO due to their considerable magnetic moments. As the size of the particles is below the coherent radius limit, in order to study the assembly in a computationally efficient form, we develop a coarse-grain macrospin model for each nanoparticle. To be more specific, we used 3 to 6 macrospins per particle. Under this level of theory, we recast the Hamiltonian in the following form:

\begin{eqnarray}
	\mathcal{H}=&-&\sum_{i,j}J_{ij}\bm{S}_i\bm{S}_j - \sum_{m,n}J_{inter}\bm{M}_m\bm{M}_n+ \sum_{m,n}\bm{M}_m[\bm{D}]\bm{M}_n \nonumber\\ 
	   &-&\sum_{i}k(cos)^2\bm{\theta}_i -g\mu_B\sum_i \bm{S}_i \bm{B}
	\label{model_assembly}
\end{eqnarray}
where $\{S_i\}$ represent the moment of macro spins within a nanoparticle while $\{M_n\}$ is the net moment of particle $n$, so that, indices $\{i,j\}$ denote summation within a nanoparticle (intra-nanoparticle interaction) while $\{m,n\}$ denote summation between nanoparticles (inter-nanoparticles interaction). The parameter $J_{ij}$ is the Heisenberg interaction coupling between macrospins  $i$ and $j$ within the nanoparticle while $J_{inter}$ describes the  interparticle interactions of the Heisenberg form.
The third term in Eq.~(\ref{model_assembly}) represents dipolar interactions between nanoparticles with $[D]$ being the dipolar tensor and it is calculated as an interaction between the net moment of each particle $\bm{M}_n$. The magnetic anisotropy constant is represented by $k$ and  $\bm{\theta}_i$ is the angle between $\bm{S}_i$ and easy axis direction while the last term represent the Zeeman contribution with $\bm{B}$ being the applied external magnetic field.    
 
 For each nanoparticle, the total magnetic moment is subdivided into $N$ macrospins, $N_1$ for the core and $N_2$ for shell, and the sum of the macrospins for the core and shell equals the total moment per nanoparticle, as determined from bulk magnetometry. Thus, the total moment per nanoparticle is divided between the core and shell contribution, separately.  For the CFO@FO core-shell nanoparticle, the core and shell moments are estimated to be $4.14 \times 10^{-20}$ Am$^2$ and $12.38 \times 10^{-20}$ Am$^2$, respectively; while for FO@CFO nanoparticle, the core and shell moments are $6.80 \times 10^{-20}$ Am$^2$ and $10.74 \times 10^{-20}$ Am$^2$, respectively. The multiple macrospins per core and shell can simulate the effects of spin canting at the core-shell and shell-vacuum interface.  The ensemble of nanoparticles is simulated by placing the macrospins for each nanoparticle on a $12 \times 12 \times 12$ grid (\textit{i.e.,} $12^3$ nanoparticles) with a mean spacing of 10 nm between nanoparticles and periodic boundary conditions.
 
 In order to define the interaction parameters, we perform a fitting procedure of the calculated single-nanoparticles energy. The atomistic energy of the system is calculated for different magnetic configurations and mapped back into the macrospin model. The tolerance of the fitting procedure ensures that all fitted parameters have an error smaller than $10^{-4} $ meV/atom. In Tables~\ref{tab:Jij} and~\ref{tab:ki}, we present the parameters used for the macrospin model. For both compounds, the indices $\{1-3\}$ refer to the core and  $\{4-6\}$ to the shell. In all cases, the anisotropy was assumed to be uniaxial for each nanoparticle, and its axis was randomly selected per particle in the assembly of nanoparticles. The interparticle interaction, $J_{inter}$, was set to $0.1$~mRy. With the aim to reduce the number of free fitting parameters, the macrospins labelled as $\{1\}$ for the core and $\{4\}$ for the shell, are imposed to interact with the same exchange interaction strength (see Table~\ref{tab:Jij}). 
 
 In Table~\ref{tab:ki}, the anisotropy contributions are distinguished between core and shell. FO region shows a considerable enhancement when it is interfaced with CFO, thus reducing the volume of the soft phase in each nanoparticle. The CFO region, being the harder magnetic phase, is the most probable nucleation region which sparks the magnetization reversal process. According to this model, the nucleation process is initiated from the surface of the CFO@FO or the core of the FO@CFO. The application of at least three macrospins per material allows us to implicity take into account by a mean magnetization state the incoherent states on the interface or surface. 
 
\begin{figure}[h]
\includegraphics[width=8.5cm]{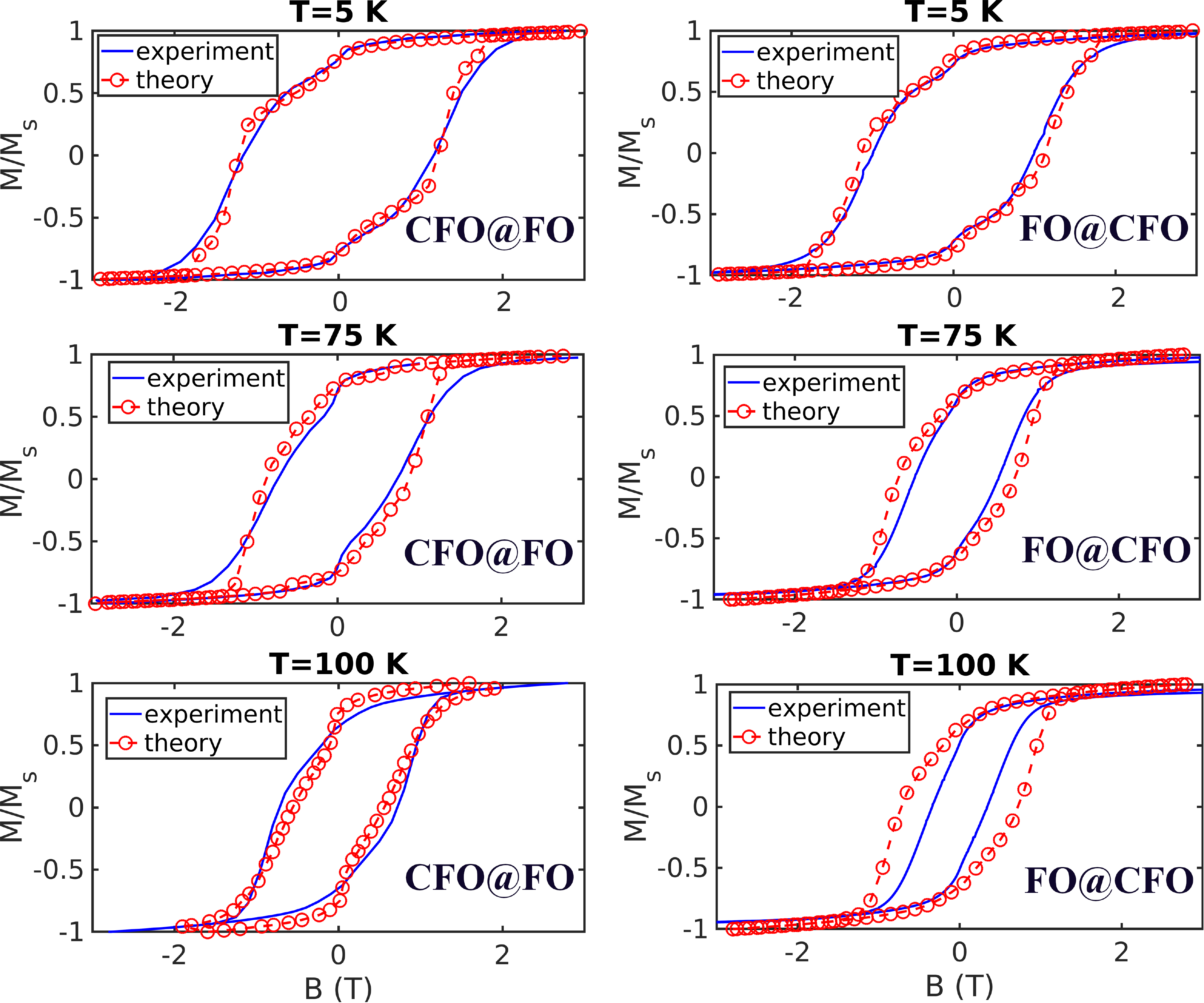}
\caption{Hysteresis loops for CFO@FO (left column) and FO@CFO (right column) assembly nanoparticles at several temperatures (5 K, 75 K and 100 K). In blue is shown the experimental hysteresis loop taken from Ref.~[\onlinecite{experiment}] for comparison with the prediction of the theoretical model. 
}
\label{cfo-fo_assv2}
\end{figure}

 Figure ~\ref{cfo-fo_assv2} shows the calculated hysteresis loops from the current macro spin model. The agreement with experimental data ~\cite{experiment} is remarkably good, especially in low temperature regime. The small value used for $J_{inter}$ indicates that even with a dense ensemble of particles, the magnetic behaviour is dominated by the intra-particle characteristics. For low temperatures, both samples are characterized by a {\textit knee} behaviour appearing right after the remanent state. Even though we have assumed a non-perfect interface, this behaviour is not so pronounced for the single nanoparticles. Thus, for the case where the nanoparticles are grouped forming an assembly, it is possible to attribute a random anisotropy axis distribution creating now a range of energy barriers to be bypassed in the cycle to complete the magnetization reversal. The introduction of dipolar interactions has an effect, that the coercive field drops down to values of about $1.1~T-1.3~T$. This value is clearly lower than the one calculated for single nanoparticles. Even though the assembly is dense, dipolar interactions do not dramatically decrease the coercive field  due to the moderate magnetic moment of the nanoparticles. It has to be noted here that even though the assembly is quite dense, we managed to reproduce the experimental data by  using a macrospin model. This fact means that we do not need to take into account the distribution of the dipolar fields because the intra-particle interactions are quite strong and, as already explained above, there cannot be any kind of magnetic domain formation in the nanoparticles. The magnetization reversal process depends strongly on the incoherent modes on the interface and surface arising from the exchange strength variations. 
 
\begin{figure}[h]
\includegraphics[width=8.5cm]{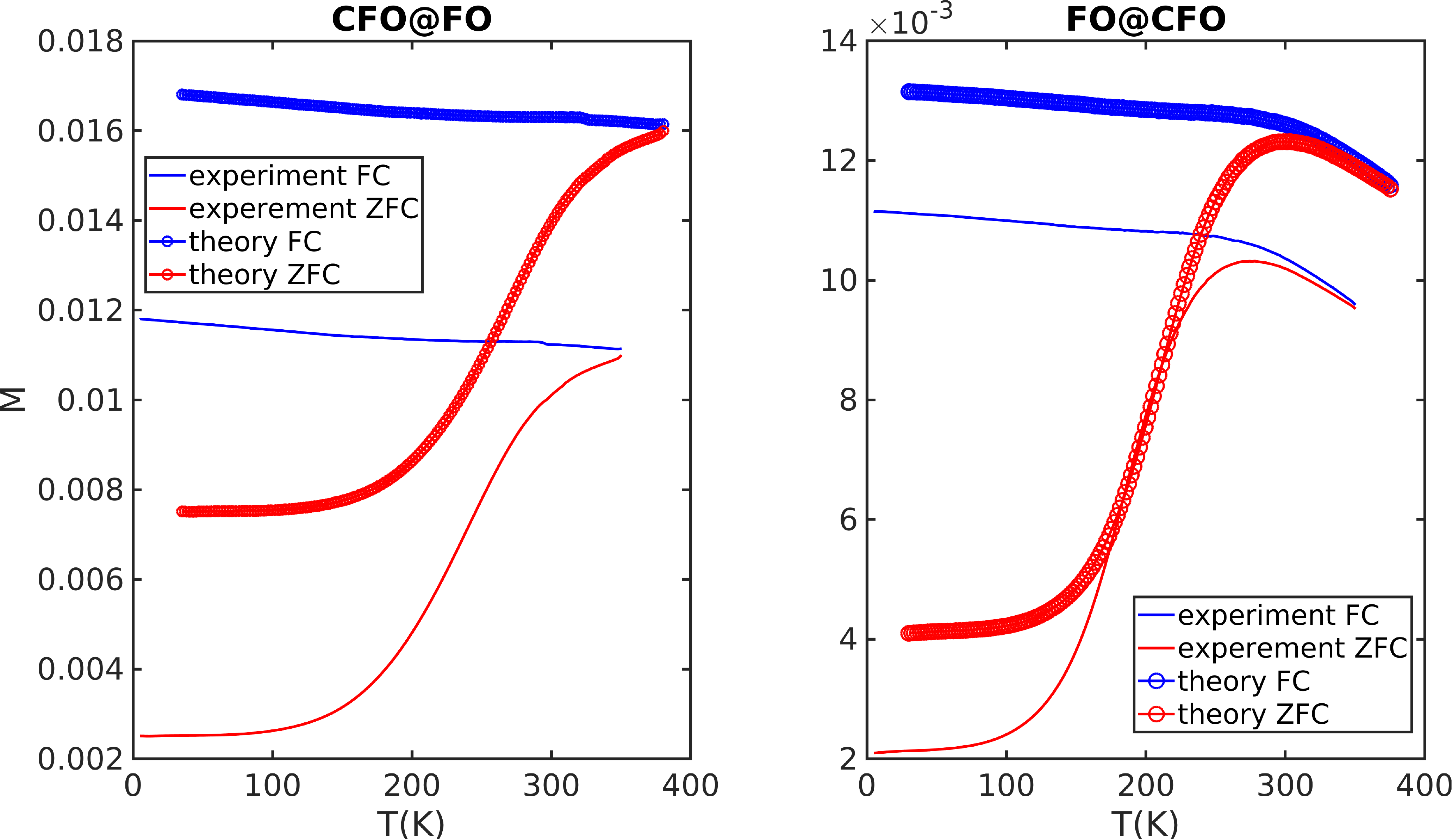}
\caption{Zero-field cooling (ZFC) and field cooling (FC) curves for the assembly nanoparticles CFO@FO (left panel) and FO@CFO (right panel). The predictions of the theoretical model (circle symbol) are compared with the experimental measurements taken from Ref.[\onlinecite{experiment}] (solid line).
}
\label{cfo-fo_assv2zfc}
\end{figure}

 The accuracy of the current model decreases as the temperature of the system is increased (cf. Fig.~\ref{cfo-fo_assv2}). For this reason, we also simulate the the ZFC/FC curves of the assemblies as shown in Fig.~\ref{cfo-fo_assv2zfc}. Notably the profile of the ZFC/FC curves is reproduced quite well with the current macrospin model however, in both cases, the blocking temperature $T_{B}$ is overestimated. The latter effect is more pronounced in the FO@CFO variant where the  $T_{B}$ is overestimated by $30$ K. This overestimation is also found in the loop simulations as the coercive field is overestimated in all cases for temperatures larger than $5$ K. The latter is a clear indication that there is a temperature dependence of the magnetic parameters, disregarded in the current model.

\begin{table}[t]
 \caption{Heisenberg interaction constant $J_{ij}$ for the macrospin model. All $J_{ij}$ are given in units of $mRy$.
 }
 \vspace{5pt}
 \centering
 \begin{ruledtabular}
 \begin{tabular}{lcccccccccc}
  Compound &$J_{12}$&$J_{13}$&$J_{23}$&$J_{24}$&$J_{34}$ &$J_{25}$&$J_{35}$&$J_{45}$&$J_{46}$&$J_{56}$\\ 
  \hline
    CFO@FO & 3.4&3.4&-2.1&2.3&1.9&2.2&1.8&2.4&2.4&-1.8   \\
    FO@CFO  & 3.1&3.1&-1.8&2.1&2.2&2.3&1.7&2.8&2.8&-1.9  \\
    
 \end{tabular}
 \end{ruledtabular}
    \label{tab:Jij}
\end{table}

\begin{table}[t]
 \caption{Anisotropy constant $k$ for the macrospin model. The units of the uniaxial magnetocrystalline anisotropy are given in $mRy$.
 }
 \vspace{5pt}
 \centering
 \begin{ruledtabular}
 \begin{tabular}{lccc|lccc}
           & &Core& &&Shell&\\
  Compound &$k_{1}$&$k_{2}$&$k_{3}$&$k_{4}$&$k_{5}$ &$k_{6}$\\ 
  \hline
    CFO@FO &  7.1 & 13.1 & 13.1 & 8.1 & 7.6 & 0.1   \\
    FO@CFO  & 0.8 & 6.4 & 6.4 & 9.2 & 8.1 & 8.0  \\
    
 \end{tabular}
 \end{ruledtabular}
    \label{tab:ki}
\end{table}

\section{Summary and Conclusions}
In conclusion, using progressively more complex models beginning with the case of isolated single nanoparticles, we have investigated the magnetic properties of ferrite nanoparticles.  The proposed macrospin model is able to describe substantially well the experimental measurements of an assembly of such core-shell nanoparticle systems. By taking the bulk values of the exchange interactions and magnetocrytalline anisotropy, we discussed the calculated hysteresis loops in terms of the ratio between the surface anisotropy with respect to the core anistropy for CFO and FO nanoparticles (cf. Fig.~\ref{cartoon} a)). By gaining some information about the internal 
parameters of isolated nanoparticles, we improved the model by considering regions with different magnetic texture, so that, the particles now have a core, interface, shell and surface (cf. Fig.~\ref{cartoon} b)). This improvement requires consideration of more types of exchange coupling between the different regions as well as magnetocrystalline anistropies per region. By scaling exchange interactions and anisotropies in the interface and surface, we can get information about the details of the hysteresis loops so that we have full control over features seen in the experimental hysteresis loops, such as, the {\it knee} present in the vicinity of the remanent field that are measured in experiments for an assembly of nanoparticles (cf.~Fig.~\ref{cfo-fo_assv2}). In the last stage of the macromagnetic model, we brought the core-shell nanoparticles, with their internal structure, all together forming a cubic crystal structure (see Fig.~\ref{cartoon} c)). This system is describe by the hamiltonian of Eq.~(\ref{model_assembly}). In doing so, the model is capable of describing with high fidelity the experimental hysteresis loops, ZFC and FC
curves. Overall, experimental and theoretical results are in close agreement although discrepancies in the hysteresis loops increase at higher temperatures. It is well-known in literature that the Heisenberg exchange interaction is a function of the temperature~\cite{attila}. We speculate here that the small deviation of the calculated hysteresis loops with respect to the experimental ones at 100 K is due to the fact we do not include a temperature dependence in the estimated Heisenberg exchange interactions. 

Overall, and based on the good description of the experimental results, the current model underpins the proposed internal structure of the nanoparticles, not only with the core and shell, but also with two thin layers of interface and surface, so that, the interface plays an important role in describing the observed {\it knee} in the experimental hysteresis loops.

\begin{acknowledgments}
 The computations were enabled by resources provided by the Swedish National Infrastructure for Computing (SNIC) at Chalmers Center for Computational Science and Engineering (C3SE), High Performance Computing Center North (HPCN), and the National Supercomputer Center (NSC) partially funded by the Swedish Research Council through grant agreement no. 2016-07213.  NN and MP acknowledge financial support from the Knut and Alice Wallenberg Foundation through grant no. 2018.0060.  HS and MHP acknowledge support (synthesis and magnetic measurements) from the US Department of Energy, Office of Basic Energy Sciences, Division of Materials Sciences and Engineering under Award No. DE-FG02-07ER46438. 
 This material is based upon work supported by the National Science Foundation under Grant No. ECCS-1952957.  DAA acknowledges the support of the USF Nexus Initiative and the Swedish Fulbright Commission.  We also thank Joshua Robles for assistance with the nanoparticle synthesis.
\end{acknowledgments}

\bibliographystyle{apsrev4-2}

\begin{thebibliography}{49}%
\makeatletter

\bibitem{par1} I. Sharifi, H. Shokrollahi and S. Amiri, Ferrite-based magnetic nanofluids used in hyperthermia applications, J. Magn. Magn. Mater. {\bf 324}, 903 (2012). 

\bibitem{par2}	S. Y. Srinivasan, K. M. Paknikar, D. Bodas and V. Gajbhiye, Applications of cobalt ferrite nanoparticles in biomedical nanotechnology, Nanomedicine {\bf 13}, 1221 (2018).

\bibitem{par3}	M. W. Mushtaq, F. Kanwal, M. Imran, N. Ameen, M. Batool, A. Batool, S. Bashir, S. M. Abbas, A. Ur Rehman, S. Riaz, S. Naseem and Z. Ullah, Synthesis of surfactant-coated cobalt ferrite nanoparticles for adsorptive removal of acid blue 45 dye, Mater. Res. Express. {\bf 5}, (2018).

\bibitem{par4}	H. Zhu, S. Zhang, Y.-X. Huang, L. Wu and S. Sun, Monodisperse $M_{x}Fe_{3-x}O_{4}$ (M = Fe, Cu, Co, Mn) Nanoparticles and their electrocatalysis for oxygen reduction reaction, Nano Lett. {\bf 13}, 2947 (2013).

\bibitem{par5}  A. Quesada, C. M. Granados, O. Lopez, A. Erokhin, S. Lottini, E. Pedrosa, J. Bollero, A. Aragon, Ana M., F. Stingaciu, M. Bertoni, C. de Julián Fernández, C. Sangregorio, J. F. Fernandez, D. Berkov, and M. Christensen, Energy Product Enhancement in Imperfectly Exchange-Coupled Nanocomposite Magnets, Adv. Elect. Mater. {\bf 2}, 1500365 (2016). 

\bibitem{par6} Ferrite nanoparticles: Synthesis, characterisation and applications in electronic device", Materials Science and Engineering: B {\bf 215}, 37 (2017). 

\bibitem{mag_NPs} M. A. G. Soler and L. G. Paterno {\it Nanostructures} (William Andrew Publishing, 2017)


\bibitem{coey} J. M. D. Coey, {\it Magnetism and Magnetic Materials} (Cambridge University Press, 2012).

\bibitem{experiment} C. Kons, K.L. Krycka, J. Robles, N. Ntallis, M. Pereiro, M.-H. Phan, H. Srikanth, J.A. Borchers and D.A. Arena, Spin Canting in Exchange Coupled Bi-Magnetic Nanoparticles: Interfacial Effects and Hard/Soft Layer Ordering, arXiv:2105.11501 [cond-mat.mes-hall].

\bibitem{cullity} B. D. Cullity and C. D. Graham {\it Introduction to magnetic Materials} (Willey, 2008)

\bibitem{an1} Y.H. Hou, Y.J. Zhao, Z.W. Liu, H.Y. Yu, X.C. Zhong, W.Q. Qiu, D.C. Zeng and L.S. Wen, Structural, electronic and magnetic properties of partially inverse spinel CoFe2O4: A first-principles study, J. Phys. D.: Appl. Phys. {\bf 43}, 445003 (2010).

\bibitem{an2} N. Daffe, F. Choueikani, S. Neveu, M. Arrio, A. Juhin, P. Ohresser, V. Dupuis, P. Sainctavit, N. Daffe,F. Choueikani, S. Neveu, M. Arrio, A. Juhin, V. Dupuis, P. Sainctavit, Magnetic anisotropies and cationic distribution in CoFe2O4 nanoparticles prepared by co-precipitation route: Influence of particle size and stoichiometry, J. Magn. Magn. Mater. {\bf 460}, 243 (2018). 


\bibitem{Jij_s} C.M. Srivastava, G. Srinivasan and N.G. Nanadikar, Exchange constants in spinel ferrites, Phys. Rev. B {\bf 19}, 1 (1979).

\bibitem{ntallis} N. Ntallis and K.G.  Efthimiadis, Size dependence of the magnetization reversal in a ferromagnetic particle, J. Magn. Magn. Mater. {\bf 99}, 373 (2015).

\bibitem{attila} A. Szilva, M. Costa, A. Bergman, L. Szunyogh, L. Nordstr\"om and O. Eriksson, Interatomic Exchange Interactions for Finite-Temperature Magnetism and Nonequilibrium Spin Dynamics, Phys. Rev. Lett. {\bf 111}, 127204 (2013).






\end{thebibliography}

\end{document}